\documentclass[aps,prl,longbibliography,superscriptaddress,reprint]{revtex4-2}

\usepackage{graphicx}
\usepackage{amsmath}
\usepackage{multirow} 
\usepackage{physics} 
\usepackage{braket}
\usepackage{times}
\usepackage{gensymb}
\usepackage{hyperref}
\hypersetup{colorlinks=true, citecolor=blue, urlcolor=blue, linkcolor=blue}

\begin{document}


\title{Experimental Demonstration of Swift Analytical Universal Control over Nearby Transitions}

\author{Yue Li}\email{Equal contribution.}
\affiliation{CAS Key Laboratory of Microscale Magnetic Resonance and School of Physical Sciences, University of Science and Technology of China, Hefei 230026, China}
\affiliation{CAS Center for Excellence in Quantum Information and Quantum Physics, University of Science and Technology of China, Hefei 230026, China}

\author{Zhi-Cheng He}  \email{Equal contribution.}  
\affiliation{Guangdong Provincial Key Laboratory of Quantum Engineering and Quantum Materials, and School of Physics\\  and Telecommunication Engineering, South China Normal University, Guangzhou 510006, China}

\author{Xinxing Yuan} \email{Equal contribution.}
\affiliation{CAS Key Laboratory of Microscale Magnetic Resonance and School of Physical Sciences, University of Science and Technology of China, Hefei 230026, China}
\affiliation{CAS Center for Excellence in Quantum Information and Quantum Physics, University of Science and Technology of China, Hefei 230026, China}

\author{Mengxiang Zhang}
\affiliation{CAS Key Laboratory of Microscale Magnetic Resonance and School of Physical Sciences, University of Science and Technology of China, Hefei 230026, China}
\affiliation{CAS Center for Excellence in Quantum Information and Quantum Physics, University of Science and Technology of China, Hefei 230026, China}

\author{Chang Liu}
\affiliation{CAS Key Laboratory of Microscale Magnetic Resonance and School of Physical Sciences, University of Science and Technology of China, Hefei 230026, China}
\affiliation{CAS Center for Excellence in Quantum Information and Quantum Physics, University of Science and Technology of China, Hefei 230026, China}

\author{Yi-Xuan Wu}  
\affiliation{Guangdong Provincial Key Laboratory of Quantum Engineering and Quantum Materials, and School of Physics\\  and Telecommunication Engineering, South China Normal University, Guangzhou 510006, China}

\author{Mingdong Zhu}
\affiliation{CAS Key Laboratory of Microscale Magnetic Resonance and School of Physical Sciences, University of Science and Technology of China, Hefei 230026, China}
\affiliation{CAS Center for Excellence in Quantum Information and Quantum Physics, University of Science and Technology of China, Hefei 230026, China}

\author{Xi Qin}
\affiliation{CAS Key Laboratory of Microscale Magnetic Resonance and School of Physical Sciences, University of Science and Technology of China, Hefei 230026, China}
\affiliation{CAS Center for Excellence in Quantum Information and Quantum Physics, University of Science and Technology of China, Hefei 230026, China}

\author{Zheng-Yuan Xue}  \email{zyxue83@163.com}
\affiliation{Guangdong Provincial Key Laboratory of Quantum Engineering and Quantum Materials, and School of Physics\\  and Telecommunication Engineering, South China Normal University, Guangzhou 510006, China}

\affiliation{Guangdong-Hong Kong Joint Laboratory of Quantum Matter, and  Frontier Research Institute for Physics,\\ South China Normal University, Guangzhou 510006, China}

\author{Yiheng Lin} \email{yiheng@ustc.edu.cn}
\affiliation{CAS Key Laboratory of Microscale Magnetic Resonance and School of Physical Sciences, University of Science and Technology of China, Hefei 230026, China}
\affiliation{CAS Center for Excellence in Quantum Information and Quantum Physics, University of Science and Technology of China, Hefei 230026, China}

\author{Jiangfeng Du} \email{djf@ustc.edu.cn}
\affiliation{CAS Key Laboratory of Microscale Magnetic Resonance and School of Physical Sciences, University of Science and Technology of China, Hefei 230026, China}
\affiliation{CAS Center for Excellence in Quantum Information and Quantum Physics, University of Science and Technology of China, Hefei 230026, China}

\begin{abstract}
Along with the scaling of dimensions in quantum systems, transitions between the system's energy levels would become close in frequency, which  are conventionally resolved by weak and lengthy pulses. Here, we extend and experimentally demonstrate analytically based swift quantum control techniques on a four-level trapped ion system, where we perform individual or simultaneous control over two pairs of spectrally nearby transitions with tailored time-varied drive, achieving operational fidelities ranging from 99.2(3)\% to 99.6(3)\%. Remarkably, we achieve approximately an order of magnitude speed up comparing with the case of weak square pulse for a general control.  Therefore, our demonstration may be beneficial to a broad range of quantum systems with crowded spectrum, for spectroscopy, quantum information processing and quantum simulation.

\end{abstract}

\date{\today}

\maketitle  
Despite frequent utilization of two-level systems for many quantum enhanced applications, most practical systems are intrinsically  of many levels nature, whose energy separations come from couplings of internal degrees of freedom or from splittings by external fields. Besides, for applications with quantum many-body system, such as, high-sensitivity precision measurements \cite{giovannetti_quantum-enhanced_2004}, large-scale quantum computation \cite{cqed} and simulation of quantum many-body system \cite{monroe_programmable_2021}, either more qubits or larger internal complexities must be involved, and then the energy or associated frequency spectrum would become crowded. In these cases, spectrally resolving a transition with an external drive field for high-fidelity manipulation would become difficult, which may limit the scalability of quantum systems, and may hinder large-scale quantum tasks where precise quantum control over individual transition is required. 

A straightforward solution for this limitation is to narrow the range of Fourier frequency components from the pulse of the drive field, for example to apply smooth-shaped pulses \cite{vandersypen_nmr_2005, ballance_high-fidelity_2016, chou2017preparation, gaebler_high-fidelity_2016, lin_quantum_2020} and using narrow linewidth transitions \cite{hayes_eliminating_2020}. However, in general cases, a desired transition rate determines the minimum temporal integral of pulse strength, i.e. ``pulse area'', and the lower field strength in turn results in longer pulse duration, leaving the control vulnerable to frequency drifts, decoherence or limiting the number of operations in given finite coherence times. Meanwhile, insufficiently weak pulse strength would create unwanted transition in the adjacent spectral line, giving rise to a population leakage or cross-talk error in the control.

To tackle with the problem of spectral crowding while keeping the operation fast, one idea is to abandon the assumption that the unwanted nearby transitions remain without being driven, but rather allowing a controlled evolution where the overall desired operation is reached at the end of the drive pulse. By numerically searching for desired pulse shape with optimal control techniques \cite{motzoi_simple_2009, forney_multifrequency_2010, gambetta_analytic_2011, motzoi_improving_2013, rebentrost_optimal_2009, rebentrost_optimal_2018}, an overall operation close enough to the target one can be obtained. However, the numerical pulse shaping can only be done in a case-by-case way, and only one certain shape can be obtained under a particular set of conditions. Besides, due to the piecewise constant control ansatz, the complex numerical shape may be inaccurate in cases of bounded bandwidth. 

On the other hand,  for two-level quantum systems, analytic solution for detuned single or multiple axis driving become available  \cite{barnes_analytically_2012, barnes_exactly_2013, economou_analytical_2015}. This brings an opportunity to better understand the underlying physics, and converting a full numerical optimization to a parametric boundary problem, where a series of solutions can be obtained when the boundary conditions are met. Thus, one has the additional freedom to incorporate various pulse-shaping techniques, and find experimental-friendly simple pulse solutions. We note that this analytic solution can be extended to drive desired transitions in multi-level quantum systems, while leaving adjacent transitions intact. Therefore, such a technique can go beyond addressing a certain transition, and can be  tailored to simultaneously drive nearby transitions. The combination of individual and simultaneous control would lead to the capability of arbitrary manipulations in the nearby transitions, and may assist in quantum logic operations on multi-level systems \cite{ Zhu2006radial, PhysRevLett.114.010501, chou2017preparation, Randall2018experimental, Low_qudit_2020, Wang_qudit_2020, Ringbauer_qudit_2021, mei2021experimental}. Remarkably, as this technique removes the weak drive limitation, it might be also useful for spectroscopy, where a faster probe to obtain more information might be possible. 

Here, we extend these analytical based swift quantum control techniques to a four-level system, where individual or simultaneous quantum control over the two pair of levels can be effectively obtained. We then experimentally demonstrate individual operation where a transition is driven while the other is intact, and simultaneous operations where quantum phase and Hadamard gate operations are performed on both transitions. We achieve operation fidelities ranging from 99.2(3)\% to 99.6(3)\%, limited by decoherence and  experimental imperfections.

\begin{figure}[tb]
	\begin{center}
		\includegraphics[width=8.5cm]{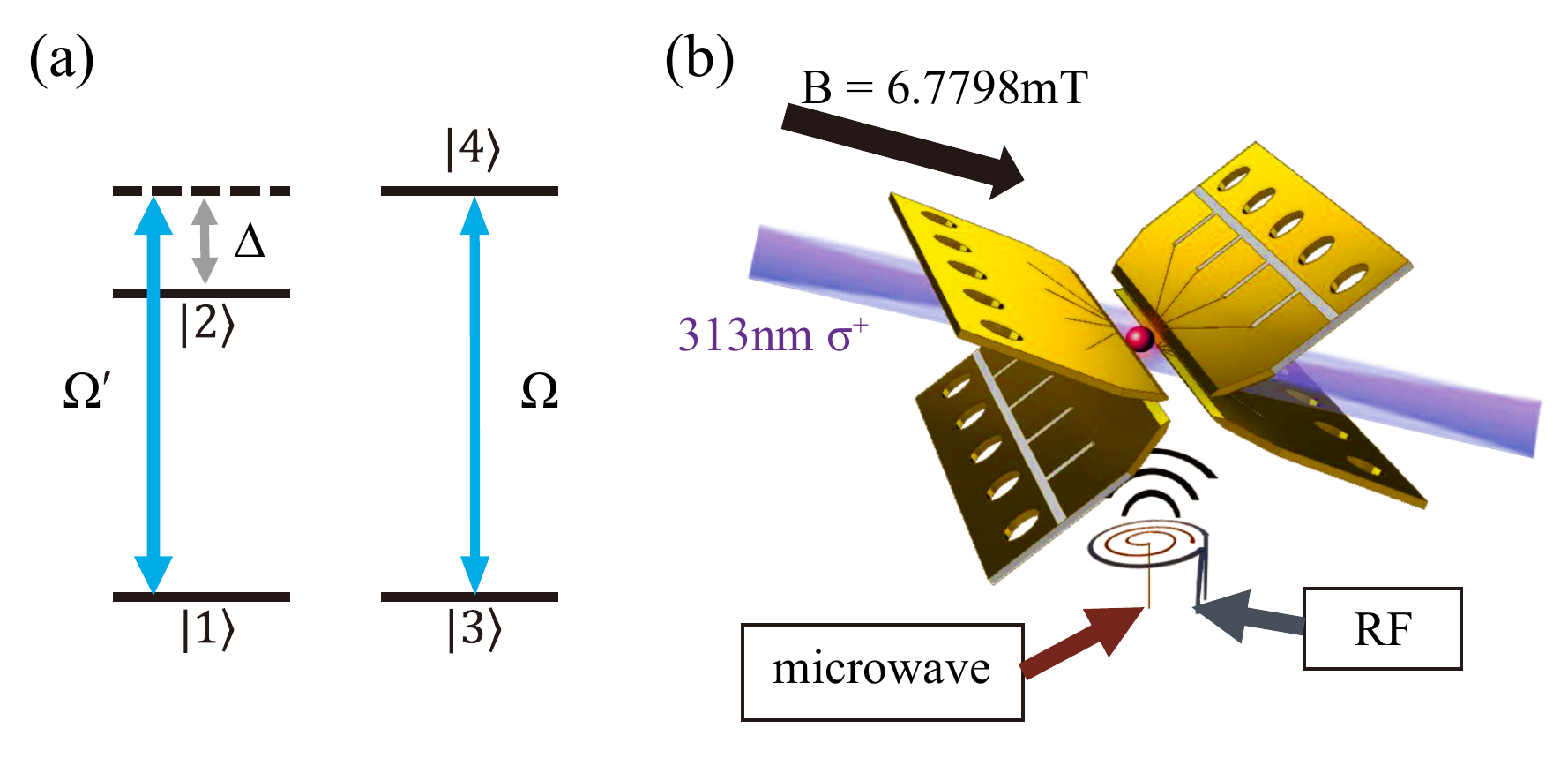}
        \caption{
        Illustration of the scheme and the experimental setup. (a) A four-level system with two pairs of closely spaced transitions is considered, where a small detuning \textbf{$\Delta$} exists between transitions with respective resonant frequency of $\omega_{12}$ and $\omega_{34}$. With a same external drive, Rabi rates for these transitions are denoted as $\Omega^\prime$ and $\Omega$ respectively.
        (b) A trapped-ion system with a $^9\rm{Be}^+$ ion trapped in a linear Paul trap. A series of controlled 313 nm laser beams with $\sigma^+$ polarization along the magnetic field direction is applied for Doppler cooling, optical pumping and resonant fluorescence detection. The magnetic field is provided by a set of $\rm{Sm}_2\rm{Co}_{17}$ permanent magnets forming a magnetic field of 6.7798 mT at the ion. Two impedance matched antennas are set under the trap, where one is for RF  drive and the other is for microwave drive.}
		\label{fig:trap}
	\end{center}
\end{figure}

We firstly extend previous two-level schemes \cite{barnes_analytically_2012, barnes_exactly_2013, economou_analytical_2015} to the case of four levels $\{ \ket{1},\ket{2}, \ket{3},\ket{4} \}$ forming two pairs of nearby transitions, as shown in Fig. \ref{fig:trap}(a), to implement a fast, high-fidelity universal control over the two transitions. Labeling the transition frequency between $\ket{i}\leftrightarrow\ket{j}$ as $\omega_{ij}$, when the transition with $\omega_{34}$ is driven resonantly by an external field with time-varied Rabi rate $\Omega$, the transition with $\omega_{12}$ is driven simultaneously by the same field with Rabi rate $\Omega^\prime$ with detuning $\Delta$ from resonance. Here we consider the case that $\Omega \sim \Omega^\prime$
and with a fixed ratio; the transition frequencies across the two pairs $\{\ket{1},\ket{2}\}\leftrightarrow\{\ket{3},\ket{4}\}$ are much larger than $\omega_{12}$ or $\omega_{34}$, thus we can limit our discussion within each pair. In the rotating framework, we obtain the interacting  Hamiltonian as 
\begin{eqnarray} \label{hami}
 H(t) = \frac{1}{2}\begin{pmatrix}
\Delta & \Omega'(t)e^{-i\varphi} & 0 & 0\\
\Omega'(t)e^{i\varphi} & -\Delta & 0 & 0\\
0 & 0 & 0 & \Omega(t) e^{-i\varphi}\\
0 & 0 & \Omega(t) e^{i\varphi} & 0\\
\end{pmatrix} 
\end{eqnarray}
where $\varphi$ is the phase of drive field. 
Thus, the nearby small detuned transition will be excited unintentionally, leading to phase and leakage errors, i.e., the so-called spectral crowding problem. The trivial way to deal with the problem is to drive the target transitions with a lower Rabi rate, i.e., max$\{\Omega, \Omega^\prime\} /\Delta \ll 1$, typically with max$\{\Omega, \Omega^\prime\} /\Delta \lesssim 0.1$ for a general control. However, lower Rabi rates lead to longer control time and less number of control pulses within the coherence times. As we analyze and experimentally demonstrate below, the time-dependent control in this work would enable individual or simultaneous control of both transitions even with max$\{\Omega, \Omega^\prime\} /\Delta \sim 1$, thus giving approximately an order of magnitude speed up.  

To illustrate the idea for swift analytic-based control, we consider the general evolution operator of Hamiltonian in Eq. (\ref{hami}) with a duration $[0,T]$ as 
\begin{eqnarray} \label{Uhami}
\mqty(\dmat[0]{U'(t), U(t)}),
\end{eqnarray}
where $U(t)$ and $U'(t)$ are the evolution operator for the resonant and detuned subspace, respectively. To obtain precise control over the transitions, we need to   obtain target dynamics for both subspaces, preferably beyond the weak drive limit. While the evolution in the resonant subspace is related to the pulse area and phase given by $\Omega(t)e^{i\varphi}$ in Eq. (\ref{hami}) in a straightforward way, the complex dynamics of the detuned subspace can be obtained, by extending Refs.  \cite{barnes_analytically_2012, barnes_exactly_2013, economou_analytical_2015},  as \cite{sm}  
\begin{eqnarray}
\label{Time Evolution Operator Non-zero}
U'(t)=U_{R}U_{C}(t)U_{R}^{\dagger}
\end{eqnarray}
where $U_{R}$ is 
\begin{eqnarray}
\label{Phase Operator}
U_{R}=\exp \left[-i\frac{\pi}{4} 
\left(
\begin{array}{cc}
0       &e^{-i(\varphi+\frac{\pi}{2})}\\
 e^{i(\varphi+\frac{\pi}{2})}      &0
\end{array}
\right)\right]
\end{eqnarray} 
and $U_{C}(t)=U_{0}(t)\cdot U_{0}^{\dagger}(0)$ with 
\begin{eqnarray}
\label{Time Evolution Operator1}
U_{0}(t)=\left(
\begin{array}{cc}
e^{i\xi_{-}(t)}\cos{\zeta(t)}       &-e^{-i(\xi_{+}(t)+\varphi)}\sin{\zeta(t)}\\
e^{i(\xi_{+}(t)+\varphi)}\sin{\zeta(t)}      &e^{-i\xi_{-}(t)}\cos{\zeta(t)}
\end{array}
\right),~~~~~
\end{eqnarray}
which is parameterized by $\zeta(t)$, and $\xi_{\pm}(t)=\frac{1}{2}\int^{t}_{0}\Delta\sqrt{1-\frac{(2\dot{\zeta}(t))^{2}}{\Delta^2}} \csc{[2\zeta(t)]}dt' \pm \frac{1}{2}\arcsin{\frac{2\dot{\zeta}(t)}{\Delta}}$. This analytic solution allows us to achieve any SU(2) matrix, and the required $\Omega^{\prime}(t)$ is determined by $\zeta(t)$ and $\Delta$ as
\begin{eqnarray}
\Omega^{\prime}(t)=\frac{2\Ddot{\zeta}(t)}{\Delta\sqrt{1-\frac{(2\dot{\zeta}(t))^{2}}{\Delta^{2}}}}- \Delta\sqrt{1-\frac{(2\dot{\zeta}(t))^{2}}{\Delta^{2}}}\cot{2\zeta(t)}.~~~~~
\label{eqrabi_rate}
\end{eqnarray}
Thus it is possible to design $\zeta(t)$ for arbitrary control in the detuned subspace. For the resonant subspace, given a fixed value of $\Omega/\Omega^\prime$, Eq. (\ref{eqrabi_rate}) leads to a waveform of $\Omega(t)$ for a controlled rotation; concatenated segments with varied $\varphi$ give arbitrary control \cite{sm}. To obtain $\zeta(t)$, we can further decompose $\zeta(t)$ into a weighted sum of analytic functions, e.g., sinusoidal functions, with a period equal to the total operation time $T$, and we can numerically find proper weights to obtain a required overall operation at $T$, under further constrains of $\dot{\zeta}(t)^{2}/\Delta^{2}<1$ and $\cot[2\zeta(t)]$ being finite. In practice, we set $T\sim {1}/{\Delta}$, then $\dot{\zeta}(t)$ and $\ddot{\zeta}(t)$ are on the order of $\Delta$ and $\Delta^{2}$, respectively. Thus in general the two terms in Eq. (\ref{eqrabi_rate}) are both on the order of $\Delta$ and are not canceled with each other,  so that the average of Rabi rates can be roughly on the order of $\Delta$, and the operation time is much shorter than the weak driven case.

In our experiment, we trap a $^9\rm{Be}^+$ ion in a linear Paul trap \cite{RevModPhys.75.281} as shown in Fig. \ref{fig:trap}(b), and the desired Hilbert space is defined within the $2s~^2S_{1/2}$ hyperfine states $\ket{F, m_F}$ of the ion, with $\ket{1} \equiv \ket{1,1}$, $\ket{2} \equiv \ket{1,0}$, $\ket{3} \equiv \ket{2,1}$ and $\ket{4} \equiv \ket{2,0}$. Applying an ambient magnetic field  at 6.7798 mT, we have $\omega_{12} = 2\pi\times 48.798$ MHz, $\omega_{34} = 2\pi\times 48.879$ MHz, $\Delta\equiv\omega_{34}-\omega_{12}=2\pi\times 81$ kHz, and $\omega_{13} = 2\pi\times1.16~\rm{GHz}$. We apply radio-frequency (RF) fields driving transitions $\ket{1}\leftrightarrow\ket{2}$ and $\ket{3}\leftrightarrow\ket{4}$. RFs are generated by an arbitrary-wave-generator (AWG) with a power amplifier \cite{sm}, and thus can be fully controlled by programming the AWG. When we drive the transition $\ket{3}\leftrightarrow\ket{4}$ on resonance with Rabi rate $\Omega$, the transition $\ket{1}\leftrightarrow\ket{2}$ is also driven with Rabi rate $\Omega^\prime$ and a detuning $\Delta$, thus implementing the Hamiltonian in Eq. (\ref{hami}). 
The Rabi rate ratio $\Omega/\Omega^\prime$ is experimentally measured to be 1.7, and remains fixed throughout the experiment since it depends on the coupling constants. To assist state preparation and detection, we also apply microwave fields to a separate antenna, as shown in Fig. \ref{fig:trap}(b), sourced by a power amplified frequency-doubled direct-digital-synthesizer.

\begin{figure}[tb]
    \centering
    \includegraphics[width=8cm]{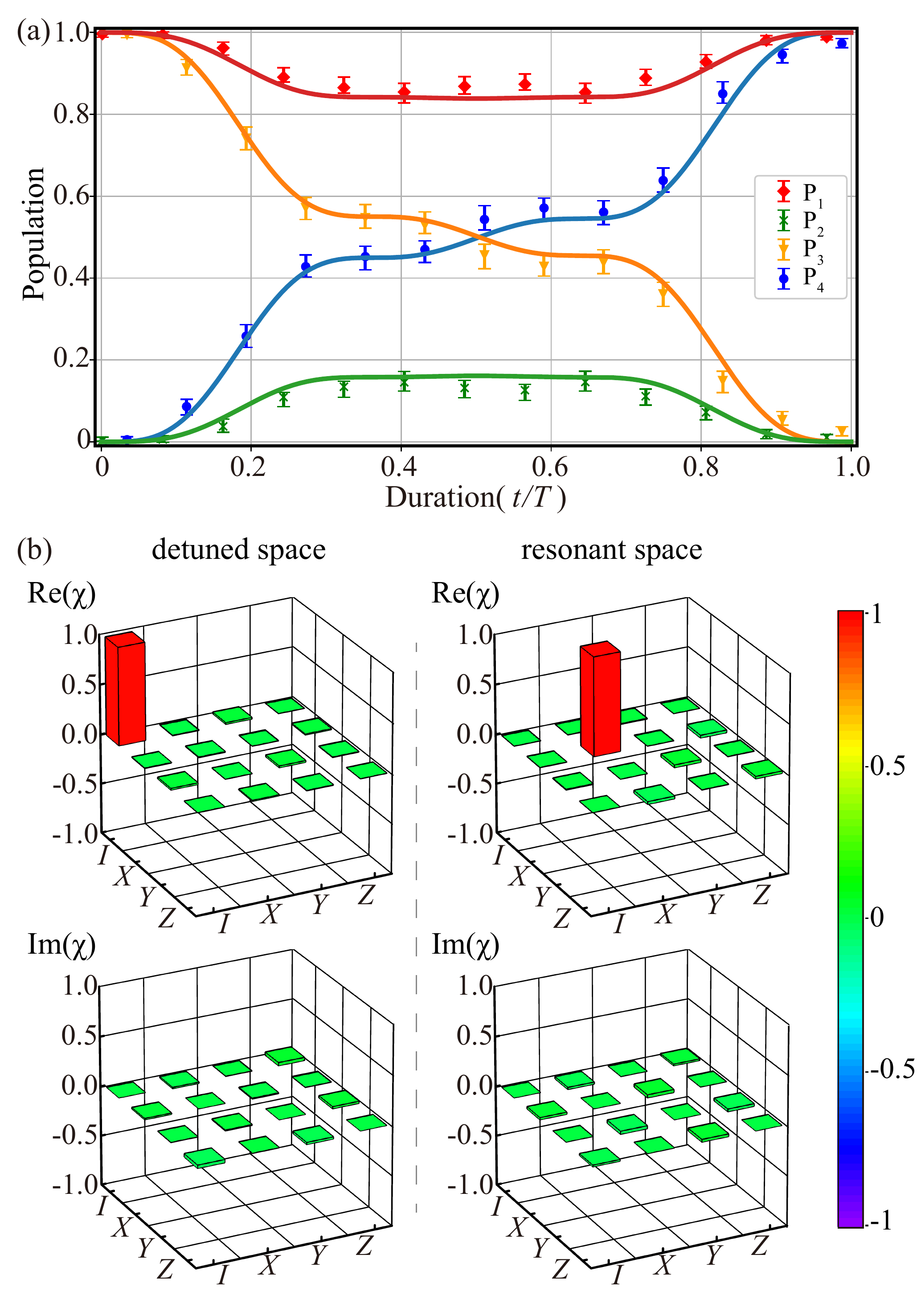}
    \caption{ Performance of individual control.
    (a) The population dynamics of $P_i$ for each basis state $\ket{i}$ under the applied control pulse. The solid lines are obtained from numerical simulation without considering experimental imperfections, overlapped with experimentally obtained populations.  Each measurement is obtained from a maximum likelihood estimation with 1000 trials. Error bars are obtained from bootstrap resampling with 95\% confidence interval \cite{keith18, gaebler_high-fidelity_2016, Clark}. (b) Process  matrices ($\chi$) for the individual control, obtained in a separate experimental run from (a). The height of the bars represents the amplitude of the matrix components, with the real and imaginary parts shown. Ideally, only the $II$ element for detuned space and $XX$ element for resonant space of the process matrix is unity and all other components are zero. Detailed analysis of the data is presented in Ref. \cite{sm}.}
    
    \label{fig:cont_e}
\end{figure} 

As a demonstration of individual control over one of the two closely spaced transitions, while leaving the other intact, we perform a simultaneous operation on both pairs of transitions with $U(T)=X$ and $U^\prime(T)=I$ ($\varphi=0$), where $X$ is a Pauli operator and $I$ is an identity operation. According to Eq. (\ref{eqrabi_rate}), we can obtain $\Omega(t)$ from $\zeta(t)$ and $\Delta$, where we set
\begin{eqnarray}
\zeta(t)=\phi_Z+\sum_{n=3}^{5} A_{n}\sin^{n}{\left(\pi\frac{t}{T}\right)},
\label{zeta}
\end{eqnarray}
with $\phi_Z=\pi/4$, leading to $\dot\zeta(0)=\dot\zeta(T)= \Ddot\zeta(0)=\Ddot\zeta(T)=0$, thus $\xi_+(0)=\xi_-(0)=0$, $\xi_+(T)=\xi_-(T)$ and $\Omega(0)=\Omega(T)=0$. When we design a specific waveform, we can set the boundary conditions and find parameters by numerical optimization with trust-region-dogleg method \cite{conntrust}. 
By setting $T =8.88$ $\mu$s for $\Delta = 2\pi\times$81 kHz and with numerical optimization, we obtain $A_{3,4,5}=\{-0.793,0.464,-0.085\}$ respectively, giving   $\xi_+(T)=\xi_-(T)=2\pi$. Thus the average of Rabi rates $\bar{\Omega}$=$2\pi\times$56 kHz and $\bar{\Omega}^\prime$ = $2\pi\times$33 kHz, comparable to $\Delta$. The quantum dynamics is shown in  Fig. \ref{fig:cont_e} (a), where the solid lines are obtained by theoretical simulation while the   points  are the experimental results. The quantum process can be described by $\rho_{out} = \sum_{j,k} E_j \rho_{in} E^{\dagger}_k \chi_{jk}$, where $\rho_{in}$ is the density matrix of the initial state, $\rho_{out}$ is the density matrix of the state after the process, $E_{j, k} \in \{I, X, Y ,Z\}$ are basis operators, $I$ is the identity operation, and $X$, $Y$ and $Z$ represent Pauli operators. Thus the matrix $\chi$ uniquely represents the process. To estimate the components of $\chi$ for the operation, we perform process tomography \cite{PhysRevLett.112.050502,Raibe_Process,Nielsen}, where we initialize the ion to various states, apply the operation, and measure the outcome. For example, to measure components of $U$, we first apply Doppler cooling, optical pumping with 313 nm laser pulses \cite{sm} to prepare $\ket{2,2}$ state, and apply resonant microwave $\pi$ or $\pi/2$ pulses to transfer to states of $\ket{3}$, $\ket{4}$, $(\ket{3}+\ket{4})/\sqrt{2}$ and $(\ket{3}+i\ket{4})/\sqrt{2}$, respectively. After applying the operation, we map one state in the pair to $\ket{1,-1}$ 
and the other to $\ket{2,2}$ by a series of microwave pulses, followed by resonant fluorescent measurement by driving $\ket{2,2}$ to $2p~^2P_{3/2}$ for 400~$\rm{\mu s}$, while the $\ket{1,-1}$ state is out of resonance and appears dark. We typically collect on average 25 counts for $\ket{2,2}$ and less than 1 count for dark states. We perform each measurement for 1000 trials to reconstruct $\rho_{out}$ by maximum likelihood state tomography \cite{keith18, gaebler_high-fidelity_2016, Clark}, and from which we further obtain the process matrix components, as shown in Fig. \ref{fig:cont_e} (b).  
Similar procedure is repeated to measure components of $U^\prime(T)$. We obtain process fidelity \cite{PhysRevLett.112.050502,Wang} $F_U = 99.5(4)\%$ and $F_{U^\prime} = 99.6(3)\%$, respectively, through the definition of 
\begin{equation}
    F=\frac{|Tr(\chi_{\rm exp}\chi_{\rm ideal})|}{\sqrt{Tr(\chi_{\rm exp}\chi_{\rm exp}^{\dagger})Tr(\chi_{\rm ideal}\chi_{\rm ideal}^{\dagger})}},
    \label{fidelity}
    \end{equation}
where $\chi_{\rm ideal}$ is the ideal process matrix for the corresponding operation and $\chi_{\rm exp}$ is the measured one. In our experiment, the  measured coherence times for  the resonant and detuned subspaces are approximately 1.9 ms and 1.1 ms, respectively,  with Ramsey type experiment, which will  contribute up to approximately 0.2\% operation infidelity.  Experimentally, we also observe fast drifts of the ambient magnetic field after each calibration, which, together with other unknown ones, we attribute as other possible error sources. 
Finally, to compare with the case of weak square pulse, we performed a numerical simulation \cite{sm}. Given a realistic decoherence of $\Delta/40$ and $\Omega/\Omega^\prime=1.7$ as in the experiment and without further imperfections, the optimal control fidelity with square pulse is reached at $\Omega/\Delta\sim 0.08$, while for our implementation a similar fidelity can be  achieved at $\bar{\Omega} /\Delta\sim 0.7$, and thus decrease the required operation time by approximately an order of magnitude.

\begin{figure}[tb]
	\begin{center}
		\includegraphics[width=8.6cm]{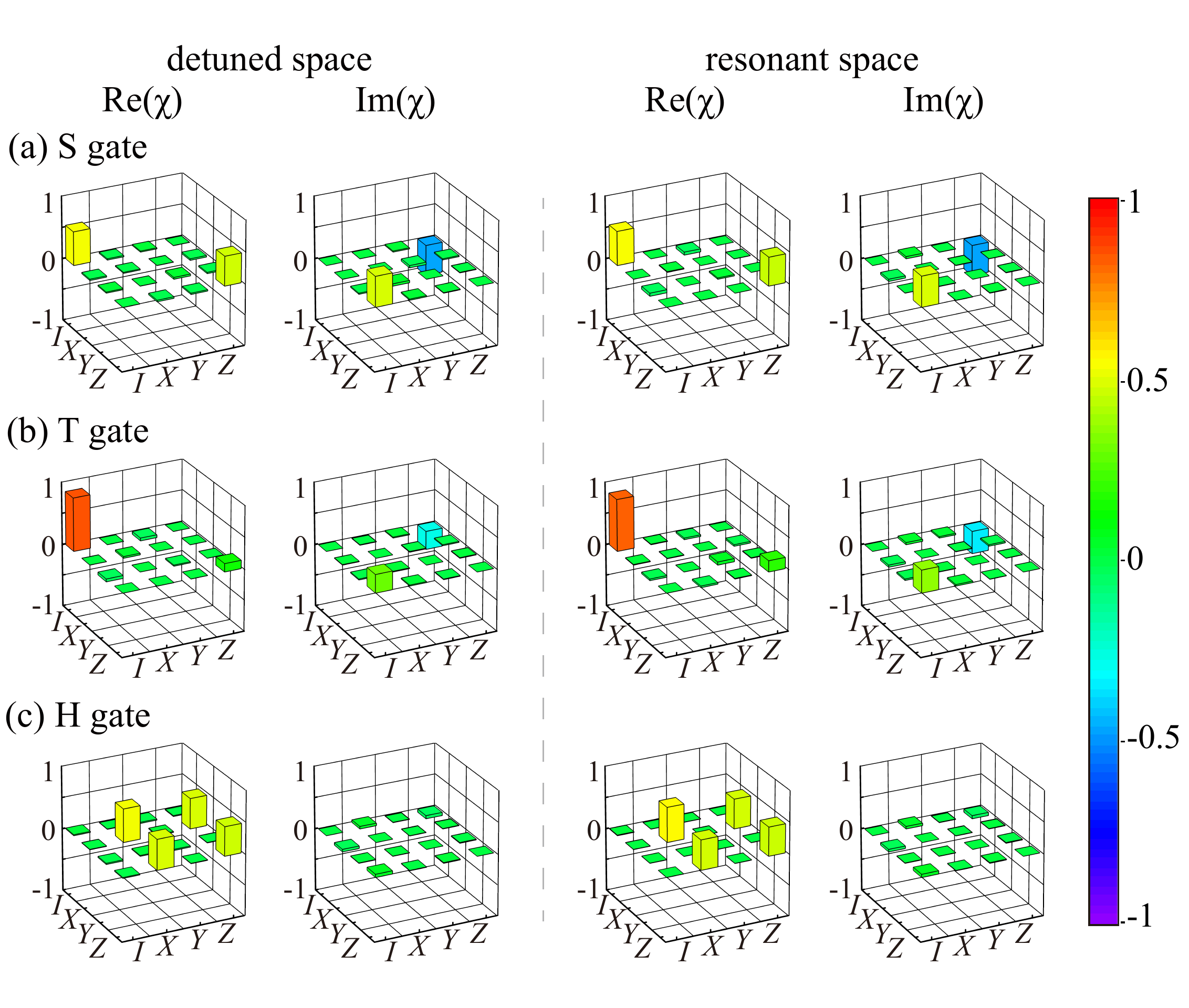}
        \caption{Bar chart for the process matrices of simultaneous control for both transitions.  Despite a small but non-negligible detuning between the two subspaces exists, a same gate can be achieved by a time-dependent drive, for (a) S gate, (b) T gate and (c) H gate, with the real and imaginary components shown. Detailed analysis of the data is presented in Ref. \cite{sm}.}
		\label{fig:single_gate}
	\end{center}
\end{figure} 

We next demonstrate simultaneous quantum control over both pairs of transitions, by implementing $U(\tau)=U^\prime(\tau)$ as quantum logic gates of S gate, T gate and Hadamard (H) gate \cite{sm}, in a two-step way with their corresponding manipulation times being $\tau_{1}$ and $\tau_{2}$, respectively, so that the total gate-time is $\tau=\tau_{1}+\tau_{2}$. 
In the  phase gate cases, a rotation $R_z(\phi)$ around the $Z$ axis with angle $\phi$ is applied in both detuned and resonant subspaces, where $\phi_S=\pi/4$, $\phi_T=\pi/8$ for S and T gates, respectively, by setting $\xi_+ ^\phi(\tau_{1,2}) = \xi_-^\phi(\tau_{1,2}) = \phi_{S, T}/2$. According to Eq. (\ref{zeta}), we can obtain specific parameters for $\Omega^\prime(t)$ in  Eq. (\ref{eqrabi_rate}). Then, we can apply two  waveforms similar to Eq. (\ref{zeta}) with different initial phase $\phi=\varphi_1-\varphi_2$ for the two steps to construct a desired phase gate, leading the first step to  $U_1^{'}(\tau_{1}) = R_z(\phi/2) $ and $U_1(\tau_{1}) = X $ with the initial phase of the pulse being $\varphi_1=0$, and for the second step   $U_2^{'}(\tau_{2}) = R_z(\phi/2)$ and $U_2(\tau_{2}) = R_{z}(\phi)X$ with the initial phase being $\varphi_2=-\phi$. In this way, we can get $U^{'}(\tau) = U(\tau) = R_z(\phi)$.  As for H gate, in the first step, we set  $\varphi_1=\pi/2$, $\phi_H = 3\pi/8$ in Eq. (\ref{zeta}) and optimize $\{A_{n}\}$ to obtain $\xi_+ ^\phi(\tau_{1}) = \xi_-^\phi(\tau_{1}) = \pi/2$, which lead to $U_1^{' H}(\tau_{1}) = H $ and $U_1^H(\tau_{1}) = \sqrt{Y}$, and for the  second step, we can apply a waveform that is the same as the individual control in Eq. (\ref{zeta}) with  $\varphi_2=0$, to get $U_2^{' H}(\tau_{2}) = I $ and $U_2^H(\tau_{2}) = X $.   The two operations form $U^{'}(\tau) = U(\tau) =  H$.   We design waveforms with gate time $\tau_{S,T,H} = \{20.77, 22.29, 25.37\}$ $\rm{\mu}$s respectively for each gate, similar to the process above and perform process tomography, with details in supplemental material \cite{sm}. The process matrices are shown in Fig. \ref{fig:single_gate}, with process fidelity $F_{S}'$= 99.4(3)\%, $F_{S}$= 99.4(3)\%, $F_{T}'$= 99.3(3)\%, $F_{T}$= 99.4(4)\%, $F_{H}'$= 99.2(2)\% and $F_{H}$=  99.2(3)\%, where  $F_{S/T/H}'$ and $F_{S/T/H}$ correspond to   S, T and H gates on the detuned and resonant transitions, respectively.The decoherence affects the measured  fidelity with a contribution up to approximate 0.3\%, 0.3\% and 0.4\% for S, T and H gates, respectively.

In conclusion, we experimentally demonstrate quantum control over two pairs of nearby transitions with time-dependent pulses, where the waveform parameters can be readily obtained with the help of an analytical model. We demonstrate individual and simultaneous control of each pair of transitions, reaching a process fidelity up to 99.6(3)\%. Our work can be extended to a range of systems. For example in recent experiments  with trapped molecular ions,  adjacent transitions can be spaced on the order of kHz, rendering the transition pulse duration on the order of ms with smooth edges \cite{chou2017preparation}; and for long ion chain, coupled motional modes can space below 10 kHz  \cite{Zhu2006radial,mei2021experimental} and superconducting transmon qubits with weak anharmonic levels \cite{PhysRevLett.114.010501}. For future research, it might be interesting to enhance the robustness for the designed shaped pulses or even incorporate optimization requirements into the process of finding a target solution \cite{soare_experimental_2014-1}. In addition, as more levels may be involved for many-body or many-degree-of-freedom quantum systems,  extending the current mathematical model to cover more levels that are closely spaced may be also of practical interest. On the other hand, our demonstration involves four levels orthogonal to each other, thus some extension would be needed when the transitions involve a shared energy level, such as $\Lambda$-, V-, and ladder-type systems. Regarding quantum sensing and metrology, this demonstration might be beneficial to vector magnetometry utilizing nitrogen-vacancy centers \cite{Wang_NV_2015} to produce a fast probe driving nearby transitions corresponding to a few crystal orientations.

\bigskip 
\begin{acknowledgments}
We thank Pengfei Wang and Xinhua Peng for helpful discussions. We acknowledge support from the National Natural Science Foundation of China (grants No. 92165206, No. 11974330 and No. 11874156), the Chinese Academy of Sciences (Grant No. XDC07000000), Anhui Initiative in Quantum Information Technologies (Grant No. AHY050000), and the Fundamental Research Funds for the Central Universities. 
\end{acknowledgments}


\section{Supplemental Material}

\section{Exact solution and speed-up comparing with square pulse}

We here expand the analytical solution in Refs. \cite{barnes_exactly_2013,economou_analytical_2015} to a general form. The Hamiltonian for a two-level quantum system that is  driven by a classical field in a detuned way is
\begin{eqnarray}
H=\frac{1}{2}\left(
\begin{array}{cc}
\Delta       &\Omega^{\prime}(t) e^{-i\varphi}\\
\Omega^{\prime}(t) e^{i\varphi}      &-\Delta
\end{array}
\right),
\end{eqnarray}
where $\Delta$ is the detuning between the transition frequency  and the driving field frequency, with $\Omega^{\prime}(t)$ and $\varphi$ being the amplitude and phase of the driving field, respectively. To get the solution, we first apply a $\pi/2$ rotation around the vertical axis of $\varphi$ axis in x-y plane, denoted as $U_{R}$, and the Hamiltonian will be changed to
\begin{eqnarray}
\label{Hamitonian_Rotated}
H_{R}=\frac{1}{2}\left(
\begin{array}{cc}
\Omega^{\prime}(t)       &-\Delta e^{-i\varphi}\\
-\Delta e^{i\varphi}      &-\Omega^{\prime}(t)
\end{array}
\right).
\end{eqnarray}
Generally, the corresponding time evolution operator for the above Hamiltonian can be written as
\begin{eqnarray}
U_{0}=\left(
\begin{array}{cc}
u_{11}(t)       &-u^{*}_{21}(t)\\
u_{21} (t)     &u^{*}_{11}(t)
\end{array}
\right),
\end{eqnarray}
with $|u_{11}(t)|^{2}+|u_{21}(t)|^{2}=1$. 

In the rotating framework of $S(t)=\exp(-i\frac{1}{2}\int^{t}_{0}\Omega^{\prime}(t')dt'\sigma_{z})$, the Schrödinger equation of the evolution operator $i\dot{U_{0}}(t)=H U_{0}(t)$ will change to
\begin{subequations}
\begin{align}
&\dot{v}_{11}(t)=i\frac{1}{2}\Delta e^{i\alpha(t)}v_{21}(t)\\
&\dot{v}_{21}(t)=i\frac{1}{2}\Delta e^{-i\alpha(t)}v_{11}(t)
\end{align}
\end{subequations}
where $$v_{11}(t)=\exp	\left(i\frac{1}{2}\int^{t}_{0} \Omega^{\prime} dt'\right) u_{11}(t),$$ $$v_{21}(t)=\exp(-i\frac{1}{2}\int^{t}_{0} \Omega^{\prime} dt') u_{21}(t),$$ and $\alpha(t)=\int^{t}_{0}\Omega^{\prime}(t')dt'-\varphi.$ 
Combining two terms, we get
\begin{eqnarray}
\frac{\dot{v_{11}}}{v_{11}}\frac{\dot{v_{21}}}{v_{21}}=-\frac{1}{4}\Delta^{2},
\end{eqnarray}
then we separate the left part as two different equation of $v_{11}$, $v_{21}$. And the right part would also be separated with a complex parameter $\kappa(t)$, i.e., $i\frac{1}{2}\Delta e^{\pm \kappa(t)}$, which respect to the Schrödinger equation from. By solving these separated differential equations, ones can get
\begin{subequations} 
\begin{align}
&v_{11}=e^{i\theta_{1}}e^{i\int^{t}_{0}\frac{1}{2}\Delta e^{\kappa(t')}dt'},\\
&v_{21}=e^{i\theta_{2}}e^{i\int^{t}_{0}\frac{1}{2}\Delta e^{-\kappa(t')}dt'},
\end{align}
\end{subequations}
here, the complex function $\kappa(t)$ is unknown yet. Here $\theta_{1}$ and $\theta_{2}$ are constant phases which are hidden in the derivation. Meanwhile, we can also denote $\alpha(t)$ in terms of $\kappa(t)$ as
\begin{eqnarray}
\label{Alpha}
\alpha(t)=-i\kappa(t)+\theta-\int^{t}_{0}\Delta\sinh{\kappa(t')}dt',
\end{eqnarray}
where $\theta=\theta_{1}-\theta_{2}$.
As $\Delta$ and $\varphi$ are real, the function $\alpha(t)$ defined by them also should be a real function, which  allows us to derive the relationship between the real and imaginary parts of $\kappa(t)$. Thus, the real and imaginary parts of $\kappa(t)$ is obtained as
\begin{subequations} 
\begin{align}
&\kappa_{R}(t)=\ln\{-\tan{[\chi(t)+C}]\},\\
&\kappa_{I}(t)=\arcsin{\frac{2\dot{\chi}(t)}{\Delta}},
\end{align}
\end{subequations}
where $\chi(t)=\frac{1}{2}\int^{t}_{0}\Delta\sin{\kappa_{I}(t^{\prime})}dt^{\prime}$, and $C$ is an integral constant. 

By applying $S^{\dagger}(t)$,  the picture will return back to the one respects to $H_{R}$, Then,   the time evolution operator can be calculated as
\begin{eqnarray}
\label{Time Evolution Operator}
U_{0}=\left(
\begin{array}{cc}
e^{i\xi_{-}(t)}\cos{\zeta(t)}       &-e^{-i(\xi_{+}(t)+\varphi)}\sin{\zeta(t)}\\
e^{i(\xi_{+}(t)+\varphi)}\sin{\zeta(t)}      &e^{-i\xi_{-}(t)}\cos{\zeta(t)}
\end{array}
\right),
\end{eqnarray}
where $\zeta(t)=\chi(t)+C$ and $\xi_{\pm}(t)=\int^{t}_{0}\frac{1}{2}\Delta\sqrt{1-\frac{(2\dot{\zeta}(t))^{2}}{\Delta^2}} \csc{(2\zeta(t))}dt' \pm \frac{1}{2}\arcsin{\frac{2\dot{\zeta}(t)}{\Delta}}.$ Here we have also set $\theta=\varphi$ for a simplicity, but in general, this constant phase can be arbitrary, as we discuss below.
Besides, we also set $\dot{\zeta}(0)=\dot{\zeta}(T)=0$ to simplify our calculation. Then the amplitude of the driving field can be obtained  as
\begin{eqnarray}
\label{Omega Form}
\Omega^{\prime}(t)=\frac{2\Ddot{\zeta}(t)}{\Delta\sqrt{1-\frac{(2\dot{\zeta}(t))^{2}}{\Delta^{2}}}}-\Delta\sqrt{1-\frac{(2\dot{\zeta}(t))^{2}}{\Delta^{2}}}\cot{2\zeta(t)}.~~~~~
\end{eqnarray}
Due to the practical restriction of the pulse shape,    the function $\zeta(t)$ should be chosen so that   $\cot{(2\zeta(t))}$ and $\sqrt{1-\frac{(2\dot{\zeta})^2}{\Delta^2}}$ are finite. Besides, to be more experimental-friendly, we also set $\Omega^{\prime}(0)=\Omega^{\prime}(T)=0$.

Furthermore, considering the boundary condition $U_{C}(0)=I$ must be met for any $C$,  we set the operator with the form of
\begin{eqnarray}
U_{C}(t)=U_{0}(t)\cdot U_{0}^{\dagger}(0).
\end{eqnarray}
Note that, the unknown constant phases $\theta_{1}$ and $\theta_{2}$ would be  canceled themselves in the calculation lead to the operator $U_{C}$ under this assumption, as been verified by our numerical simulation and experiment.
Finally, turning back to original frame, the analytical solution of the time evolution operator is
\begin{eqnarray}
\label{Time Evolution Operator Non-zero}
U^{\prime}(t)=U_{R}U_{C}(t)U_{R}^{\dagger}.
\end{eqnarray}

\begin{figure}[tb]
	\begin{center}
		\includegraphics[width=8.6cm]{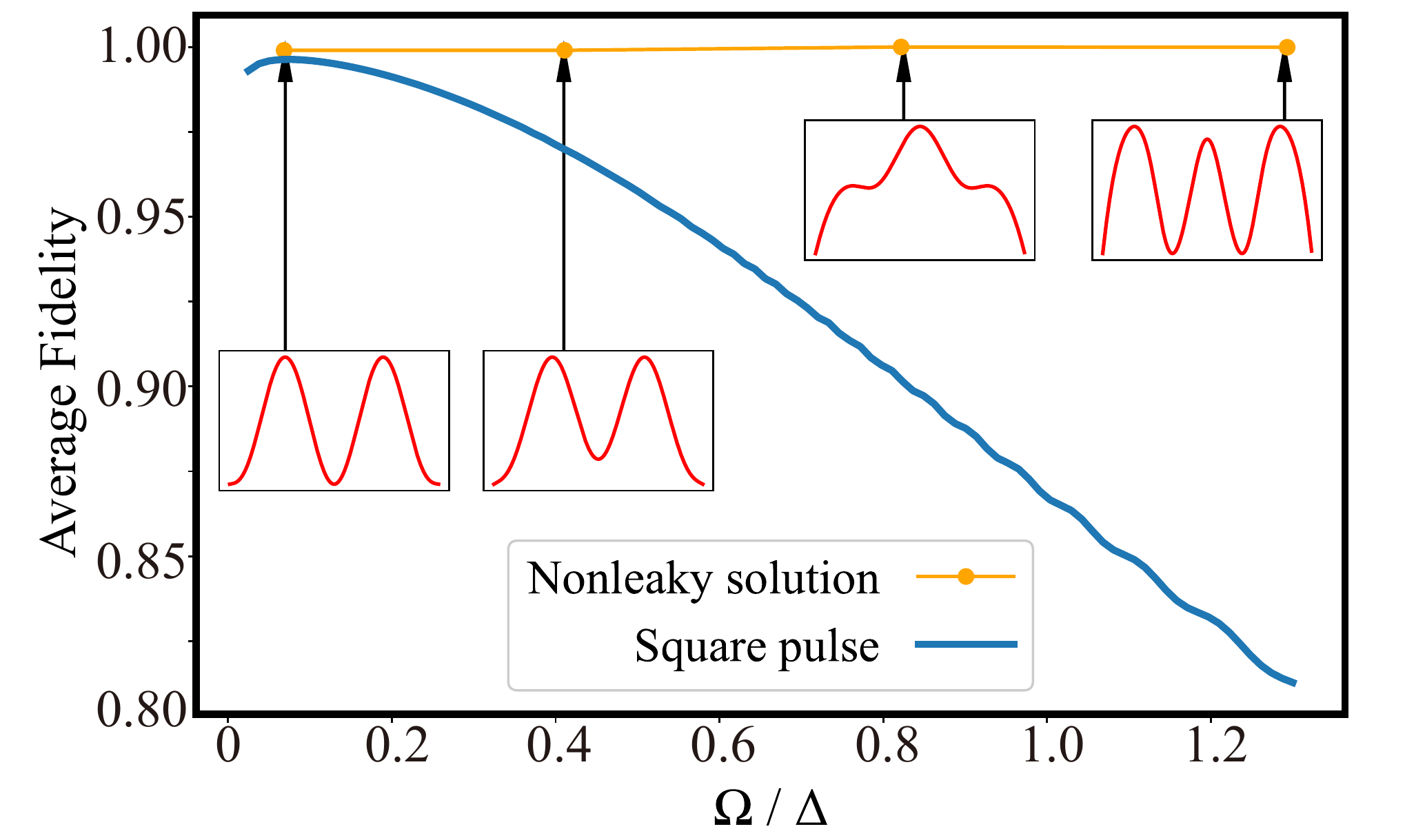}
        \caption{Simulation of average gate fidelity with dephasing rate to be $\Delta/40$, and $\Omega/\Omega^\prime=1.7$, to match with realistic experimental condition. Here we compare our analytical solution based pulse with the direct square pulse driving. The analytical solution can keep good fidelity($>99.8\%$) with $\Omega/\Delta$ increasing while the direct driving can only be good at the weak limit. The inset figures are typical waveforms at different ratio.  
		}
		\label{fig:leakage}
	\end{center}
\end{figure}

{We did numerical simulations to compare the performance between using swift analytical control and using square pulses, aiming to conduct the individual control}. The Hamiltonian of our system is 
\begin{eqnarray} \label{hami}
 H(t) = \frac{1}{2}\begin{pmatrix}
\Delta & \Omega'(t)e^{-i\varphi} & 0 & 0\\
\Omega'(t)e^{i\varphi} & -\Delta & 0 & 0\\
0 & 0 & 0 & \Omega(t) e^{-i\varphi}\\
0 & 0 & \Omega(t) e^{i\varphi} & 0\\
\end{pmatrix}.~~~~~~
\end{eqnarray}
We can design the Rabi rate $\Omega(t)$ so that $\int_0^T \Omega(t) dt = \pi$, where   $T$ is the operation duration, then we can get a $\pi$ rotation on the resonant subspace, without any restriction on the pulse shape. Meanwhile, the nearby small detuned transition can be driven simultaneously, by tuning the pulse shape to meet the form in Eq. (\ref{Omega Form}), we can get any target operator in this subspace. We simulate the effect of the unintentional drive as shown in Fig. \ref{fig:leakage}. We define the operator fidelity between the unintentional drive and $I$, where $I$ is the identity operation. We did the simulation with the dephasing rate  being $\Delta/40$ and $\Omega/\Omega^\prime=1.7$. Our analytical solution based pulse can keep good  fidelity($>99.8$\%) even with increasing $\Omega / \Delta$.

\section{Universal quantum control}

\begin{figure}[tb]
    \centering
    \includegraphics[width=8.6cm]{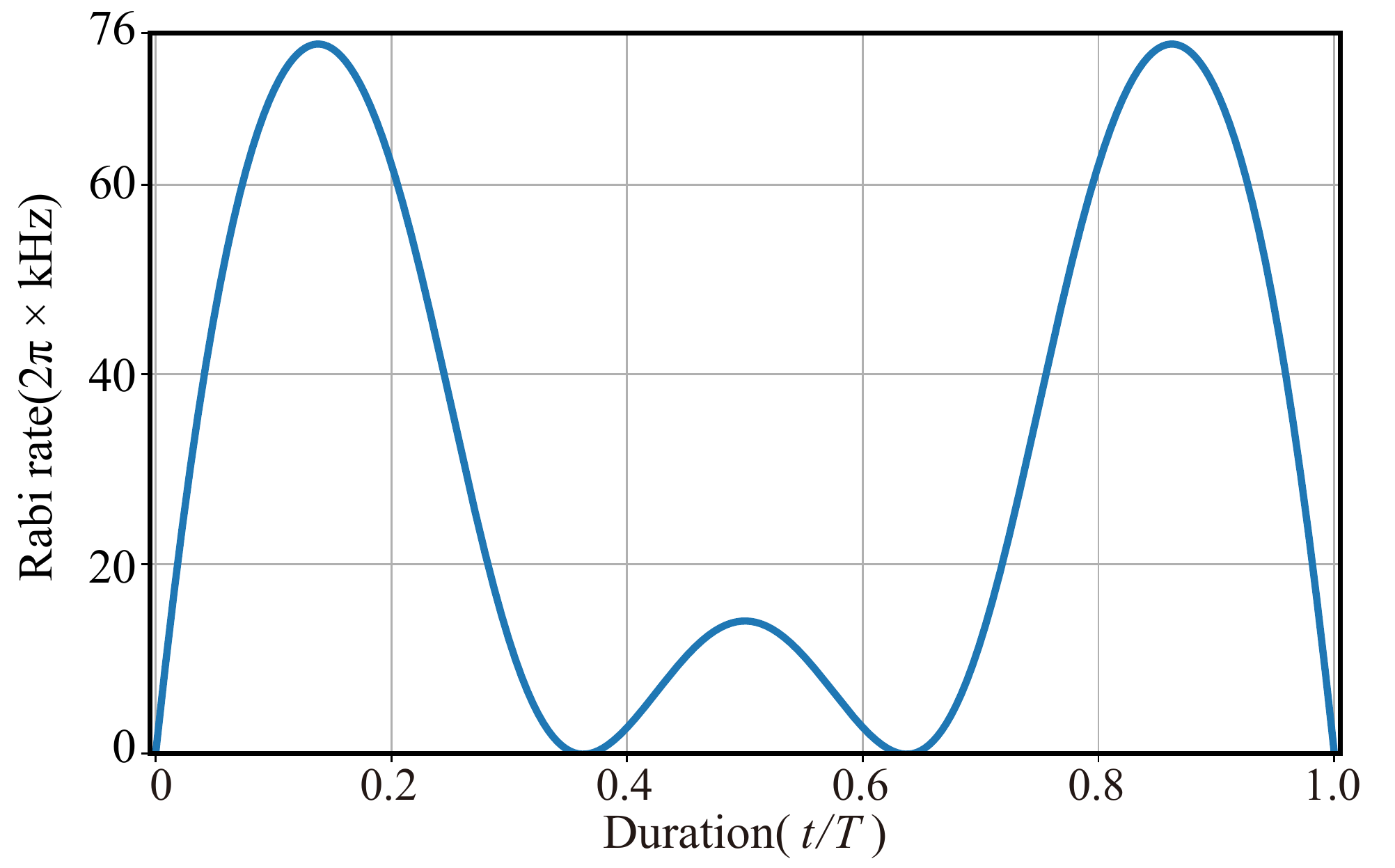}
    \caption{A typical waveform designed for individual control with parameter $A_3 = -0.793, A_4 = 0.464, A_5 = -0.085$ and gate time $T = 8.88$ $\mu$s.}
    \label{fig:waveform}
\end{figure}

Firstly, we focus on individual control (ID) over one of the two closely spaced transitions. For this purpose, we can perform an  target operation on the resonant  subspace, e.g., $U(T) = X$, while maintain the integrity of  the detuned subspace, i.e.,  $U^\prime(T) = I$, where $X$ is a Pauli operator and $I$ is an identity operation. Then, the exemplified  operator is
\[
U_{ID} = 
\renewcommand\arraystretch{1}
\begin{array}{@{}r@{}c@{}c@{}c@{}c@{}l@{}}
   &\ket{1} & \ket{2} & \ket{3} & \ket{4}   \\
   \left.\begin{array}
    {c} \ket{1} \\\ket{2} \\\ket{3} \\\ket{4} \end{array}\right(
                & \begin{array}{c} 1 \\ 0  \\ 0 \\  0 \end{array}
                & \begin{array}{c} 0 \\ 1 \\ 0 \\ 0 \end{array}
                & \begin{array}{c} 0 \\ 0 \\ 0\\ 1 \end{array}
                & \begin{array}{c} 0 \\ 0 \\ 1 \\ 0 \end{array}
                & \left)\begin{array}{c} \\ \\ \\ \\ \end{array}\right.
  \end{array}.
\]
To realize this operation, in the resonant subspace, a simple $\pi$ pulse is enough. And, in the detuned subspace, the analytic solution in Eq. (\ref{Time Evolution Operator Non-zero}) shows that the evolution operator can be written as
\[ U_{det}(t) = 
\begin{pmatrix}
e^{i\xi(t)}    &0\\
0      &e^{-i\xi(t)}
\end{pmatrix},\]
where we have set $\zeta(0)=\zeta(T)=\frac{\pi}{4}$, and assumed $\dot{\zeta}(0)=\dot{\zeta}(T)=0$ for simple calculation. These settings lead to $$\xi=\frac{1}{2}\int^{T}_{0}\Delta\sqrt{1-\frac{(2\dot{\zeta})^{2}}{\Delta^2}} \csc{(2\zeta)}dt.$$ And the function $\zeta(t)$ in experiment is set as
\begin{eqnarray}
\zeta(t)=\frac{\pi}{4}+\sum_{n=3}^{5} A_{n}\sin^{n}{(\pi\frac{t}{T})},
\end{eqnarray}
where $A_{n}$ are arbitrary parameters to be determined by various purpose.  Here, we choose $A_{1}=0$ to satisfy $\dot{\zeta}(0)=\dot{\zeta}(T)=0$ and $A_{2}=0$ to make sure $\Omega(0)=\Omega(T)=0$. The rest of $A_{n}$ could be used to determine the value of $\xi$ numerically, providing different possible gate series. Here, we choose $\xi=2\pi$ and $A_3 = -0.793, A_4 = 0.464, A_5 = -0.085$ to implement the target  individual control with short operational time, and the resulted pulse shape is shown in Fig. \ref{fig:waveform}.

In principle, arbitrary control over each subspaces can be obtained. Therefore, besides the individual control, we also present how to implement simultaneously  quantum gates, such as S gate, T gate and Hadamard (H) gate on both  subspaces. For a qubit system, these gates form a universal set of single-bit gates
In the case   S gate, it denotes a $\frac{\pi}{4}$ rotation around $\sigma_z$ axis in both detuned and resonant subspace,i.e.
\[ U^{\prime}(\tau) = 
\begin{pmatrix}
e^{i\frac{\pi}{4}} & 0\\
0 & e^{-i\frac{\pi}{4}}
\end{pmatrix}, U(\tau) = 
\begin{pmatrix}
e^{i\frac{\pi}{4}} & 0\\
0 & e^{-i\frac{\pi}{4}}
\end{pmatrix}.\]

To implement this single-bit gate, we need two $\pi$ pulse with different  $\varphi$. In the resonant subspace, we set
\begin{subequations}
\begin{align}
&\int^{\tau_{1}}_{0}\frac{1}{2}\Omega (t)dt=\frac{\pi}{2}, \quad\varphi_{1}=0,\\
&\int^{\tau_{2}}_{\tau_{1}}\frac{1}{2}\Omega (t)dt=\frac{\pi}{2}, \quad\varphi_{2}=-\frac{\pi}{4}.
\end{align}
\end{subequations}
So the gate $U(\tau)$ was implemented.
For detuned subspace, recall the analytic solution, we find the phase $\xi$ in evolution operator is independent on  $\varphi$ of the drive field, thus the phase $\xi$ in these two pulses are actually equal, and the final operator is
\[  
\begin{pmatrix}
e^{i2\xi}    &0\\
0      &e^{-i2\xi}
\end{pmatrix},\]
When $2\xi=\frac{\pi}{4}$, ones implement the gate $U'(\tau)$, so the S gates in two subspace is  obtained. The amplitudes for both steps of the waveform designed for S gate are the same with $A_{3,4,5}=\{-0.259,-0.059,-0.093\}$ and $\tau_1 = \tau_2 = 10.39~\mu$s. As for T gates, just set $2\xi'=\frac{\pi}{8}$, it can be implemented in a similar way.  The waveform parameters designed for T gate are $A_{3,4,5}=\{-0.134,-0.077,-0.197\}$ and $\tau_1 = \tau_2 = 11.15~\mu$s.

We now consider the Hadamard (H) gate for both subspaces, that is,
\[ U^{\prime}(\tau) = 
\begin{pmatrix}
1 & 1\\
1 & -1
\end{pmatrix}, U(\tau) = 
\begin{pmatrix}
1 & 1\\
1 & -1
\end{pmatrix},\]
To implement the gate in resonant subspace, we decompose H gate as $H=X\cdot\sqrt{Y}$, where $X(Y)$ denote the  Pauli matrix  $\sigma_{x}(\sigma_{y})$, which could be obtained by a resonant pulse with different phase $\varphi=0(\frac{\pi}{2})$ with the pulse area being  $\pi(\frac{\pi}{2})$. Thus, we set
\begin{subequations}
\begin{align}
&\int^{\tau_{1}}_{0}\frac{1}{2}\Omega (t)dt=\frac{\pi}{4}, \quad\varphi_{1}=\frac{\pi}{2},\\
&\int^{\tau_{2}}_{\tau_{1}}\frac{1}{2}\Omega (t)dt=\frac{\pi}{2}, \quad\varphi_{2}=0.
\end{align}
\end{subequations}
For the detuned subspace, we implement an identity  in the first part of evolution, and  an H gate for the second part. The corresponding restriction of parameter in second part is $\zeta(\tau_{1})=\zeta(\tau_{2})=3\pi/8$, and the second part of the waveform is designed with parameters $A_{3,4,5}=\{-1.093, 0.747, -0.360\}$ and $\tau_2 =16.49~\mu$s. Similar as before, the H  gate could be implemented with accurately calculated $A_{n}$ with $\xi= \pi/2$.

\section{Experimental result}
The data of $\chi$ matrix for individual or simultaneous control is shown as follows, in each of the subspaces.  $\chi_{ID/S/T/H}$ correspond to the individual control, simultaneous S, T and H gates process matrix for resonant space; $\chi^{\prime}_{ID/S/T/H}$ is the process matrix for detuned space, with $Re(\cdot)$ and $Im(\cdot)$ representing the real and imaginary components respectively:
\begin{align}
    &Re(\chi^\prime_{ID}) = \begin{pmatrix}
    0.997 && -0.008 && 0.019 && 0.005\\
    -0.010 && -0.010 && -0.005 && 0.013\\
    0.019 && -0.005 && 0.020 && -0.008\\
    0.005 && 0.013 && -0.008 && 0.004\\
    \end{pmatrix}, 
\end{align}
\begin{align}    
    &Im(\chi^\prime_{ID}) = \begin{pmatrix}
    0.000 && 0.016 && -0.05 && -0.036\\
    -0.016 && 0.000 && 0.012 && 0.005\\
    0.005 && -0.012 && 0.000 && -0.024\\
    0.036 && -0.005 && 0.024 && 0.000\\
    \end{pmatrix},
\end{align}
\begin{align}    
    &Re(\chi_{ID}) = \begin{pmatrix}
    -0.040 && 0.006 && 0.011 && 0.004\\
    0.006 && 0.995 && -0.004 && -0.030\\
    0.011 && -0.004 && 0.030 && 0.006\\
    0.004 && -0.030 && 0.006 && 0.027\\
    \end{pmatrix}, 
\end{align}
\begin{align}    
    &Im(\chi_{ID}) = \begin{pmatrix}
    0.000 && 0.024 && 0.002 && -0.019\\
    -0.024 && 0.000 && -0.033 && -0.002\\
    -0.002 && 0.034 && 0.000 && 0.030\\
    0.020 && 0.002 && -0.030 && 0.000\\
    \end{pmatrix},
\end{align}
\begin{align}
    &Re(\chi^\prime_{S}) = \begin{pmatrix}
    0.547 && -0.003 && -0.049 && 0.000\\
    -0.000 && 0.012 && -0.000 && -0.002\\
    -0.049 && -0.000 && 0.020 && -0.000\\
    0.000 && -0.002 && -0.000 && 0.460\\
    \end{pmatrix}, 
\end{align}
\begin{align}
    &Im(\chi^\prime_{S}) = \begin{pmatrix}
    0.000 && -0.035 && 0.003 && 0.491\\
    0.035 && 0.000 && -0.006 && -0.003\\
    -0.003 && 0.006 && 0.000 && 0.008\\
    -0.491 && 0.003 && -0.007 && 0.000\\
    \end{pmatrix},
\end{align}
\begin{align}    
    &Re(\chi_{S}) = \begin{pmatrix}
    0.540 && -0.020 && 0.017 && -0.003\\
    -0.020 && -0.016 && 0.003 && -0.025\\
    0.017 && 0.003 && 0.030 && -0.020\\
    -0.003 && -0.025 && -0.020 && 0.475\\
    \end{pmatrix},
\end{align}
\begin{align}    
    &Im(\chi_{S}) = \begin{pmatrix}
    0.000 && -0.001 && -0.018 && 0.487\\
    0.001 && 0.000 && 0.037 && 0.018\\
    0.018 && -0.037 && 0.000 && 0.008\\
    -0.487 && -0.018 && -0.008 && 0.000\\
    \end{pmatrix},
\end{align}
\begin{align}
    &Re(\chi^\prime_{T}) = \begin{pmatrix}
    0.832 && 0.009 && -0.026 && -0.001\\
    0.009 && -0.007 && 0.001 && -0.025\\
    -0.026 && 0.001 && 0.051 && 0.009\\
    -0.001 && -0.025 && 0.009 && 0.180\\
    \end{pmatrix}, 
\end{align}
\begin{align}    
    &Im(\chi^\prime_{T}) = \begin{pmatrix}
    0.000 && -0.035 && -0.023 && 0.363\\
    0.035 && 0.000 && 0.006 && 0.023\\
    0.023 && -0.006 && 0.000 && -0.004\\
    -0.363 && -0.023 && 0.004 && 0.000\\
    \end{pmatrix}, 
\end{align}
\begin{align}      
    &Re(\chi_{T}) = \begin{pmatrix}
    0.857 && 0.002 && -0.046 && -0.001\\
    0.002 && 0.027 && 0.001 && -0.002\\
    -0.046 && 0.001 && -0.017 && 0.002\\
    -0.001 && -0.002 && 0.002 && 0.134\\
    \end{pmatrix}, 
\end{align}
\begin{align}      
    &Im(\chi_{T}) = \begin{pmatrix}
    0.000 && 0.008 && -0.015 && 0.294\\
    -0.008 && 0.000 && 0.012 && 0.015\\
    0.015 && -0.012 && 0.000 && -0.010\\
    -0.294 && -0.015 && 0.010 && 0.000\\
    \end{pmatrix},
\end{align}
\begin{align}
    &Re(\chi^\prime_{H}) = \begin{pmatrix}
    0.010 && -0.003 && 0.004 && 0.002\\
    -0.003 && 0.561 && -0.002 && 0.480\\
    0.004 && -0.002 && 0.002 && -0.003\\
    0.002 && 0.480 && -0.003 && 0.464\\
    \end{pmatrix}, 
\end{align}
\begin{align}      
    &Im(\chi^\prime_{H}) = \begin{pmatrix}
    0.000 && -0.040 && 0.006 && 0.051\\
    0.040 && 0.000 && 0.004 && -0.006\\
    -0.006 && -0.004&& 0.000 && 0.001\\
    -0.051 && 0.006 && -0.012 && 0.000\\
    \end{pmatrix},
\end{align}
\begin{align}    
    &Re(\chi_{H}) = \begin{pmatrix}
    0.016 && -0.006 && -0.018 && -0.014\\
    -0.006 && 0.536 && 0.014 && 0.489\\
    -0.018 && 0.014 && -0.0172 && -0.006\\
    -0.014 && 0.489 && -0.006 && 0.470\\
    \end{pmatrix},
\end{align}
\begin{align}      
    &Im(\chi_{H}) = \begin{pmatrix}
    0.000 && 0.008 && 0.018 && 0.069\\
    -0.008 && 0.000 && -0.010 && -0.018\\
    -0.018 && 0.010 && 0.000 && 0.007\\
    -0.069 && 0.018 && -0.007 && 0.000\\
    \end{pmatrix}.
\end{align}

\section{Details of Experiment hardware}
We use commercial 313 nm laser sources from Precilasers Co. Ltd. (YEFL-FHG-313-0.3-CW), frequency locked to reference spctral lines of molecular iodine with a home-built saturation spectroscopy setup via the 626 nm monitor output. We use commercial AWG from Ciqtek (AWG4100). The AWG is triggered and synchronized with home-developed board and user interface (based on and modified from \cite{Langer2006}), which control direct-digital-synthesizer and Transistor-transistor logic for other pulses and record photon counts during fluorescent detection. The power amplifiers used for RF and microwave fields are Mini Circuits ZHL-100W-GAN+ and  ZHL-30W-252-S+ respectively. The maximum power we use for RF and microwave fields are 50 W and 30 W respectively. We set two separate antennas for RF drive and microwave drive, about 2 cm away from the ion. The RF antenna has a shape of a coil with around 3 cm diameter made from a single turn of copper wire. A 100 nH inductor and a 30 pF capacitor in parallel with the coil improve the coupling of the radio frequency to the antenna. The microwave field is delivered through a twin spiral antenna with a diameter of around 3 cm.

%

\end{document}